%% file: cluster-formation-sonic-arxiv.tex
\documentclass[aps,pre,twocolumn,floatfix]{revtex4-1}

\usepackage{amsmath}  % needed for \tfrac, \bmatrix, etc.
\usepackage{amsfonts} % needed for bold Greek, Fraktur, and blackboard bold
\usepackage{graphicx} % needed for figures
\usepackage{epstopdf}
\usepackage{gensymb}

\begin{document}
\title{Cluster formation via sonic depletion forces in levitated granular matter}

\author{Melody X. Lim, Anton Souslov, Vincenzo Vitelli, Heinrich M. Jaeger}
\affiliation{Department of Physics and James Franck Institute, The University of Chicago, 5720 S. Ellis Ave, Chicago, Illinois 60637, USA}
\begin{abstract}
The properties of small clusters can differ dramatically from the bulk phases of the same constituents. In equilibrium, cluster assembly has been recently explored, whereas out of equilibrium, the physical
principles of clustering remain
elusive. These principles underlie phenomena
from molecular assembly to the formation
of planets from granular matter. Here,
we introduce acoustic levitation as a platform to experimentally probe the formation of nonequilibrium small structures in a controlled
environment. We focus on the minimal models
of cluster formation: six and seven millimetre-scale particles in two dimensions.
Experiments and modelling reveal that, in contrast to thermal colloids, in non-equilibrium granular ensembles the magnitude of active fluctuations controls not only the assembly rates but also their assembly pathways
and ground-state statistics. 
These results open up new possibilities for non-invasively manipulating macroscopic particles, tuning their interactions, and directing their assembly.
\end{abstract}
\maketitle

Mechanically agitated granular matter often serves as a prototype for exploring the rich physics associated with hard sphere systems, with an effective temperature introduced by vibrating or shaking~\cite{Olafsen1998,Danna2003,Feitosa2004,Keys2007,Komatsu2015,Workamp2018}. While depletion interactions drive clustering and assembly in colloids~\cite{Manoharan2003,Sacanna2010,Meng2010,Kraft2012}, no equivalent short-range attractions exist between macroscopic grains. 
To overcome this limitation, we introduce an experimental approach based on acoustic levitation and trapping of granular matter~\cite{Gorkov1961,Wang2017,Lee2018}. 
Scattered sound establishes short-range attractions between small particles~\cite{Settnes2012}, while detuning the acoustic trap generates active fluctuations~\cite{Rudnick1990}. 
To illuminate the interplay between attractions and fluctuations, we investigate transitions among ground states of two-dimensional clusters composed of a few particles. 

In two dimensions, particle clusters with five or fewer constituents have only one compact configuration, i.e., one isostatic ground state~\cite{Perry2015} (Fig. 1a). 
However, beginning with six particles, there are an increasing number of energetically degenerate, but geometrically distinct, ground-state configurations. 
This complex energy landscape has been studied with colloids in thermal equilibrium ~\cite{Meng2010,Perry2015}. 
Here, we explore the ground-state statistics in ensembles of macroscopic particles driven by active fluctuations that emerge from the dynamics of a driven system rather than from coupling to a heat bath. Furthermore, we demonstrate how energetic degeneracies, assembly rates, and pathways are altered during out-of-equilibrium assembly.

\begin{figure}
\includegraphics[width=\columnwidth]{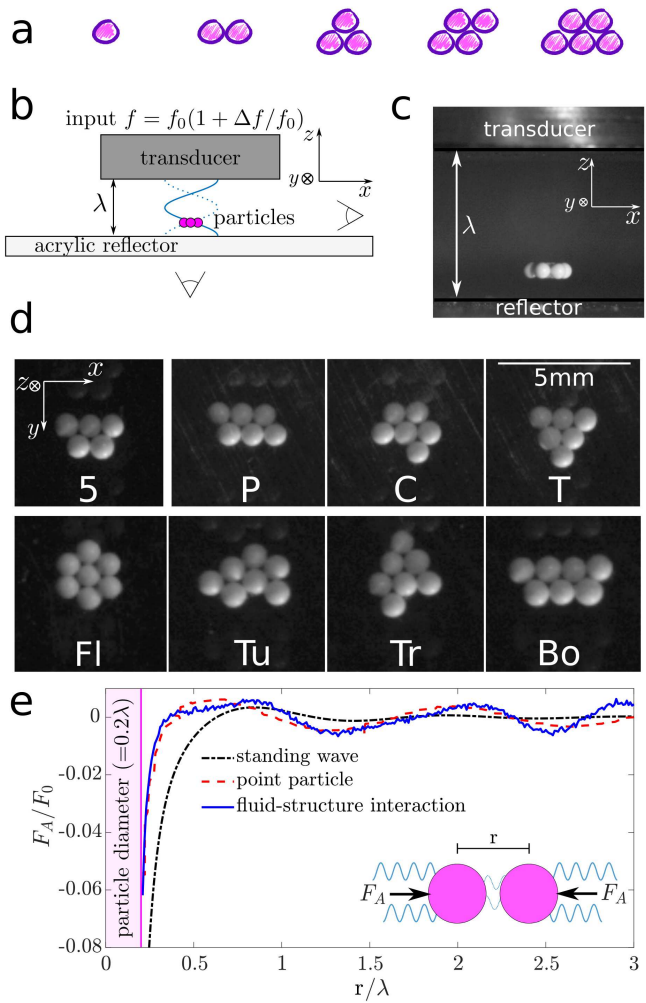}
\caption{\textbf{Acoustic levitation as a platform to assemble and manipulate clusters composed of macroscopic particles.} (a) Sketches of compact cluster configurations (isostatic ground states) for one to five particles. (b) Schematic of experimental setup. An ultrasound transducer generates sound waves in air, with speed of sound~$c_s=343$m/s. The distance between transducer and transparent acrylic reflector is chosen to create a pressure standing wave (blue line) with two nodes, at frequency~$f_0=45.651$kHz and wavelength~$c_s/f_0$. Polyethylene particles are acoustically levitated in the lower of the two nodes. Clusters are imaged from the side (b), as well as from below via a mirror (c). (c) Different cluster configurations, imaged from below. (Top) In two dimensions, there is only one 5-particle cluster configuration, but six particles can form one of three distinct ground states: parallelogram P, chevron C, and triangle T. (Bottom) Seven-particle clusters have four compact configurations: flower Fl, turtle Tu, tree Tr, and boat Bo. (d) The acoustic field generates short-range attractions within the levitation plane, which stabilises particle clusters. Force between two particles as a function of distance~$r$ between their centres, normalised by the particle weight ~$F_0 \equiv m_0 g$. Finite-element simulations (red dashed, blue solid lines) are compared to an analytical solution for particles in a standing wave~\cite{Silva2014} (black dashed-dotted line). (Inset) Schematic illustrating the sonic depletion force between two particles in an acoustic field.}
\label{fig:setup}
\end{figure}

To eliminate frictional interactions with container walls we levitate particles in a sound pressure field. 
The same field also induces short-range, tunable attractions that we here call sonic depletion forces. 
Much like depletion forces in colloids~\cite{Asakura1954} or other Casimir-like forces, these acoustically-mediated attractions can generate robust particle clusters. 
This differs from the formation of clusters in granular media due  to external confinement ~\cite{Kudrolli1997,Olafsen1998} or, transiently, due to inelastic collisions ~\cite{Goldhirsch1993,Kudrolli1997,Olafsen1998,Brilliantov2004}. 
Furthermore, the sonic depletion forces scale with the sound pressure amplitude, which enables precise control over cluster energetics. Such control provides advantages over cohesive forces due to capillary bridges, van der Waals interactions, or charging~\cite{Royer2009,Lee2015}.  
Finally, in contrast to induced electric or magnetic dipole forces~\cite{Lumay2008}, the acoustic interactions are not aligned with an applied vector field. 

Our setup is illustrated in Fig.~\ref{fig:setup}b.  We generate a standing wave of the acoustic pressure field between an ultrasound transducer and the (transparent) acrylic reflector. Polyethylene particles (diameter 710-850$\mu$m) levitate within a horizontal plane one-quarter of the gap height from the reflector.
We image these acoustically trapped particles from the side (Fig.~\ref{fig:setup}c), or from below (Fig.~\ref{fig:setup}d) using a high-speed camera.
When multiple particles are placed in the trap, they form compact clusters. Images of the resulting configurations for six- and seven-particle clusters are shown in Fig.~\ref{fig:setup}d.
Six-particle systems have three distinct ground-state configurations: parallelogram (P), chevron (C), and triangle (T). 
For 7-particle clusters, there are four distinct topologies: Flower (Fl), Tree (Tr), Turtle (Tu), and Boat (Bo)~\cite{Klein2018}. 

Whereas colloidal clusters can be stabilised by depletion forces, acoustically levitated clusters are stabilised by sonic depletion forces, 
which are short-range attractions generated by acoustic scattering.
At close approach, these Casimir-like forces $F$ between spherical particles scale as
\begin{equation}
F \sim \frac{E_0 a^6 \lambda^{-3}}{r^4}
\label{eq:casimir}
\end{equation}
where $E_0 \equiv \rho_0 v_0^2/2$ is the energy density of the sound field having amplitude $v_0$ and wavelength $\lambda$ in air (density $\rho_0$)~\cite{Silva2014}. The particles have radius $a$
and are distance $r$ ($\ll \lambda$) apart.
For arbitrary separation, these forces can be computed analytically in the absence of trap geometry~\cite{Silva2014}, or with finite-element simulations using either the Gor'kov approximation~\cite{Gorkov1961} or fluid-structure interactions~\cite{GlynneJones2013} within the trap geometry (see Methods). 
These calculations, shown in Fig.~\ref{fig:setup}e, indicate that cluster energetics are dominated by the strong short-range attractions between nearest neighbours, as captured in Eq.~(\ref{eq:casimir}).

The acoustic trap can also induce non-conservative forces. Specifically, we use the fact that the particle dynamics in the acoustic field are underdamped (in contrast to colloids in a liquid) to drive instabilities that generate active fluctuations.  
As Ref.~\cite{Rudnick1990} shows, a sound wave with frequency~$f$ tuned just slightly larger than the standing wave resonance condition acts on a levitated object with a destabilising force proportional to the object's speed.
As a result, the clusters fluctuate up and down in the trap, occasionally hitting the reflector. This impact transfers kinetic energy from center-of-mass motion to modes that bend the cluster out of its planar, two-dimensional configuration. 
For sufficiently high amplitudes, these active fluctuations can lead to rearrangements between the different ground states (see Supplementary Movies 1 \& 2). 

\begin{figure*}
\includegraphics[width=2\columnwidth]{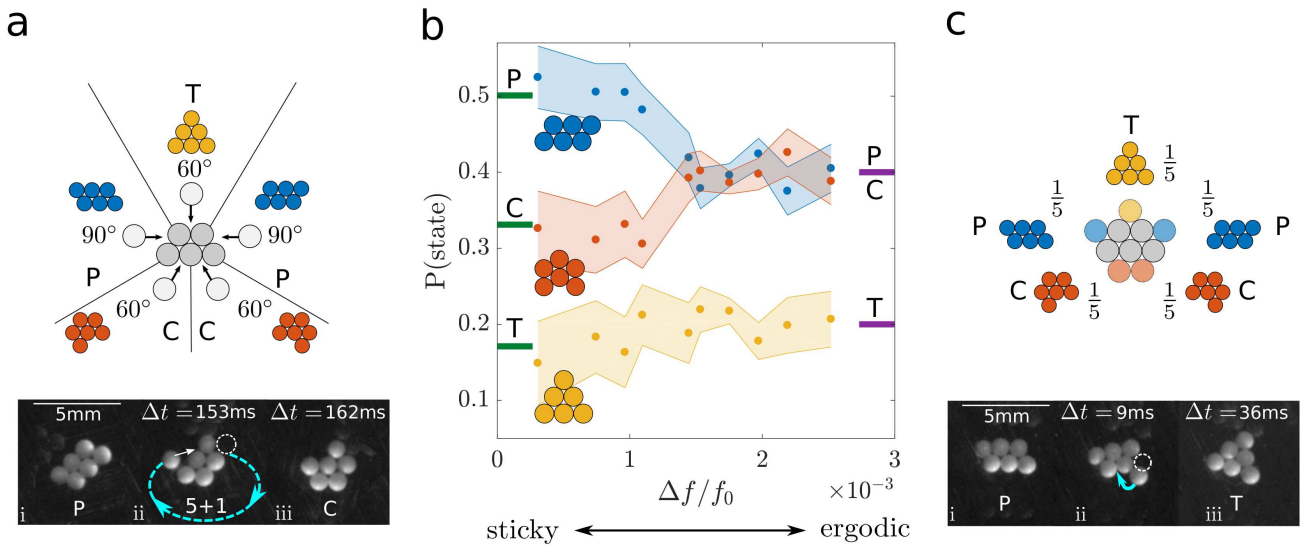}
\caption{\textbf{Tuning 6-particle assembly between sticky and ergodic limits.} 
Near resonance, cluster statistics follow sticky assembly (a).
As the acoustic trap is detuned by increasing the sound frequency (b), cluster statistics change to ergodic assembly (c).
(a) Sticky assembly: In the regime of small detuning parameter, $\Delta f/f_0 > 0$, the likelihood of a cluster configuration is determined by the geometric angle of approach of a sixth particle to the 5-particle cluster (top). (Bottom) Sequence of images from below showing a sticky rearrangement pathway. See Supplementary Movies 1 and 3 for dynamics. (b) Steady-state probabilities for 6-particle-cluster ground-state configurations as a function of detuning parameter. Standard error is indicated by shaded region. Horizontal bars indicate model predictions for the sticky and ergodic limits (see text). (c) Ergodic assembly: For larger detuning, the sixth particle has equal probability of occupying each of the five binding sites on the five-particle cluster (top).  (Bottom) Sequence of images from below showing a transition between ground states in the ergodic regime through a hinge motion. See Supplementary Movies 1 and 4 for dynamics.}
\label{fig:statistics}
\end{figure*}

Close to resonance, 6-particle clusters rearrange by ejecting a single particle, which then travels many particle diameters in a curved trajectory before it re-joins the 5-particle cluster from a random angle of approach. 
Once the particle re-joins, it becomes stuck due to the short-range attraction. This sticky, far-from-equilibrium assembly pathway is shown in Fig.~\ref{fig:statistics}a. 
The corresponding cluster statistics retain memory of the formation process~\cite{Wang2017}: the ground-state configuration is determined by the spatial angle of approach that the sixth particle takes towards the 5-particle cluster (see Supplementary Movie 3). 
Assuming that docking onto the 5-cluster is equally likely for any angle of approach (see Fig.~\ref{fig:statistics}a), the probabilities of forming P, C, or T 6-clusters are 1/2, 1/3, and 1/6, respectively, in close agreement with the data for the sticky limit (Fig.~\ref{fig:statistics}b).

By contrast, deep into the off-resonant regime, clusters rearrange by moving particles randomly along their periphery (Fig.~\ref{fig:statistics}c). 
This occurs either by single particle ejection with much shorter trajectories (i.e., no more than one particle diameter) or by `floppy' hinge motions: When all but one of the bonds to nearest neighbours is broken by active cluster fluctuations, the remaining bond acts as a flexible hinge. 
This enables the particle to swing around to a new position without leaving the cluster.  
In this off-resonant regime, we find that P and C clusters occur with equal probability and twice as often as T clusters (Fig.~\ref{fig:statistics}b). 
Such cluster statistics correspond to an unbiased sampling of configuration space, where we simply count the number of ways a 6-cluster can be formed by adding one more particle to a 5-cluster. 
This ergodic limit is indistinguishable from the thermal case, which Ref.~\cite{Perry2015} observed using 6-particle clusters composed of micrometre-sized Brownian colloids.  

By changing the ultrasound frequency, we can control the amplitude of active fluctuations and thus control the cluster rearrangement processes.
Figure~\ref{fig:statistics}b shows statistics for relative ground-state probabilities as a function of detuning parameter~$\Delta f/f_0$, where~$f_0$ ($=45.651$kHz) is the trap resonant frequency, $f$ is the driving frequency, and~$\Delta f \equiv f-f_0 > 0$. 
As the trap is detuned, cluster statistics transition smoothly from sticky to ergodic. At the same time, clusters increasingly rearrange via hinge motions (see Supplementary Movie 4).

\begin{figure*}
\includegraphics[width=2\columnwidth]{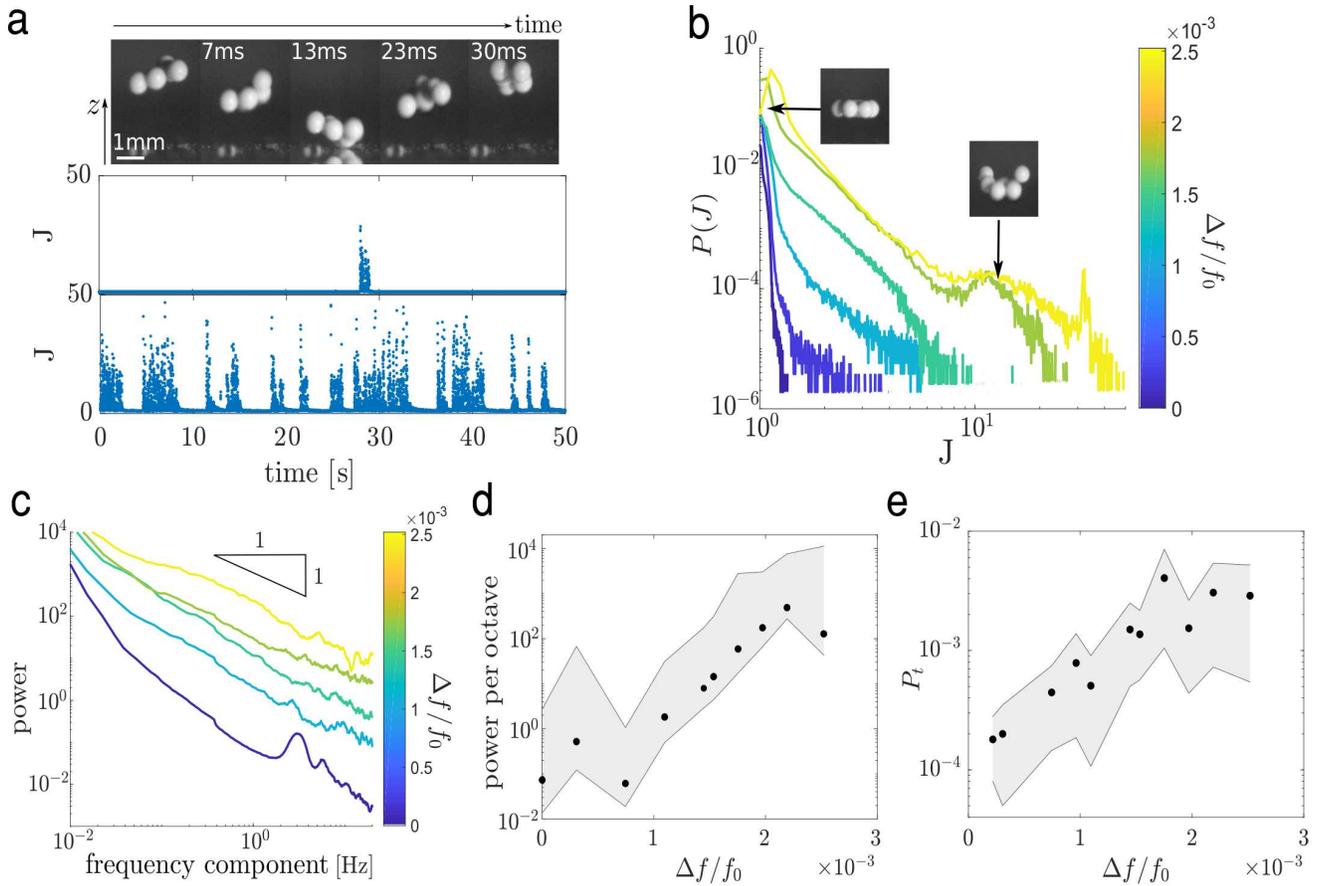}
\caption{\textbf{Out-of-plane fluctuations as a measure of effective temperature.} (a) (Top) Sequence of side images showing a cluster colliding with the reflector. See Supplementary Movie 2 for dynamics. Time series for second moment~$J$ of vertical coordinate $z$ (see Methods), for small detuning (middle) and large detuning (bottom). (b) Probability distribution of~$J$ as function of detuning parameter, obtained from time series as in part (a). Illustrative side images of clusters are shown at their value of~$J$. (c) Power spectrum of the $J$ time series. (d) Average power per octave as function of detuning parameter. (e) $P_t$, the probability of transition between any cluster ground state, as a function of detuning parameter. Shaded areas in (d) and (e) indicate the standard error.
Note the similar trends in (d) and (e): the fluctuations increase exponentially as a result of the effective temperature induced by detuning away from resonance.}
\label{fig:temperature}
\end{figure*}

The emergence of hinge motions is closely linked to out-of-plane bending, which like particle ejection is triggered by impacts against the reflector, as shown in Fig.~\ref{fig:temperature}a (see also Supplementary Movie 2). 
We quantify the associated deviation from planar configuration by computing the second moment~$J$ of the vertical pixel coordinates~$z$ associated with a cluster in side-view (see Methods).
For a fully planar configuration, $J$ is at a minimum;
if the cluster is bent out of plane, $J$ increases.
Representative time series of~$J$ for small and large detuning parameters are shown in Fig.~\ref{fig:temperature}a. 
From longer versions of such time series, the probability distributions $P(J)$ for finding a particular magnitude $J$ can be extracted. As Fig.~\ref{fig:temperature}b shows, clusters remain effectively rigid and planar for small~$\Delta f/f_0$, while further detuning generates a rapidly increasing probability of exciting large-$J$ values associated with shape-changing, out-of-plane bending fluctuations. These fluctuations also become more frequent (Fig.~\ref{fig:temperature}a, bottom), resulting in broad power spectra whose magnitude quickly rises with~$\Delta f/f_0$, while their overall character changes little (Fig.~\ref{fig:temperature}c).

When we plot the average power per octave associated with shape-changing fluctuations we find it to increase exponentially with detuning parameter~$\Delta f/f_0$ (Fig.~\ref{fig:temperature}d).
At the same time, we find that also the probability~$P_t$ of observing a transition between any two 6-particle ground states increases exponentially (Fig.~\ref{fig:temperature}e). 
Together, this shows that~$\Delta f/f_0$ plays the role of an effective temperature in an activated process: detuning the trap generates instabilities that temporarily break particle-particle bonds and allow for cluster rearrangement. 

Here, a surprising aspect is that detuning not only controls the rate, but also the type of rearrangement process. 
From Fig.~\ref{fig:statistics}b, we see that these processes have important consequences for the likelihood of observing specific ground state configurations. 
In particular, the degeneracy between parallelogram (P) and chevron (C) in the ergodic limit can be broken by moving to the regime dominated by sticky assembly. 

Driven by active fluctuations, these clusters explore an athermal ensemble. The cluster reconfigurations are instances of a general transition process through intermediate states. 
We model this process with a discrete-time Markov chain, in which state transition matrices represent the creation of specific ground state configurations through adding or removing one particle. 
To represent the various ground-state probabilities $P_i$ for a general $N$-particle cluster, we list them as $i$ components of a vector $\mathbf{P}_N$. Specifically for $N$ = 6, $\mathbf{P}_6 = (P_P, P_C, P_T)$, where the subscripts refer to the three possible configurations.
The $(i,j)$-th element of the transition matrix~$\mathrm{T}_N$ represents the probability of creating the $i$-th $N$-particle ground state by adding a single particle to the $j$-th $(N - 1)$-particle ground state. 
Similarly, the $(i,j)$-th element of the matrix $\mathrm{Q}_{N}$ captures how the $i$-th $N$-particle state is obtained by destroying the $j$-th ground state of the $(N + 1)$-particle cluster.
Under steady-state conditions, $\mathbf{P}_N$ is related to the probabilities $\mathbf{P}_{N-1}$ and $\mathbf{P}_{N+1}$ through
 \begin{align}
 \mathbf{P}_N=\mathrm{T}_{N} \mathbf{P}_{N-1}+\mathrm{Q}_{N} \mathbf{P}_{N+1}\,.
 \label{eq:general}
 \end{align}

Once $\mathrm{T}_N$ and $\mathrm{Q}_N$ are known, Eq.~(\ref{eq:general}) can be solved recursively for $\mathbf{P}_N$ (see Methods). 
For the case discussed so far, with six particles in the trap, Eq.~(\ref{eq:general}) leads to $\mathbf{P}_6=\mathrm{T}_{6} \mathbf{P}_{5}$ and $\mathbf{P}_5=\mathrm{Q}_{5} \mathbf{P}_{6}$, which gives $\mathbf{P}_6=\mathrm{T}_{6} \mathrm{Q}_{5} \mathbf{P}_{6}$.
Since removing any particle from a 6-cluster results in the same 5-cluster (so that $\mathbf{P}_{5}$ = 1), 
we have 
$\mathrm{Q}_{5}$ = $\begin{pmatrix}
1&1&1
\end{pmatrix}$.
However, the $3\times 1$ matrix $\mathrm{T}_6$ depends on whether the creation process is sticky or ergodic, i.e., its components are the docking probabilities indicated in the top panels of Fig.~\ref{fig:statistics}a, c. 
Solving for $\mathbf{P}_6$ then gives the values indicated by the horizontal bars along either side of Fig.~\ref{fig:statistics}b, in close agreement with the data. 

\begin{figure*}
\includegraphics[width=2\columnwidth]{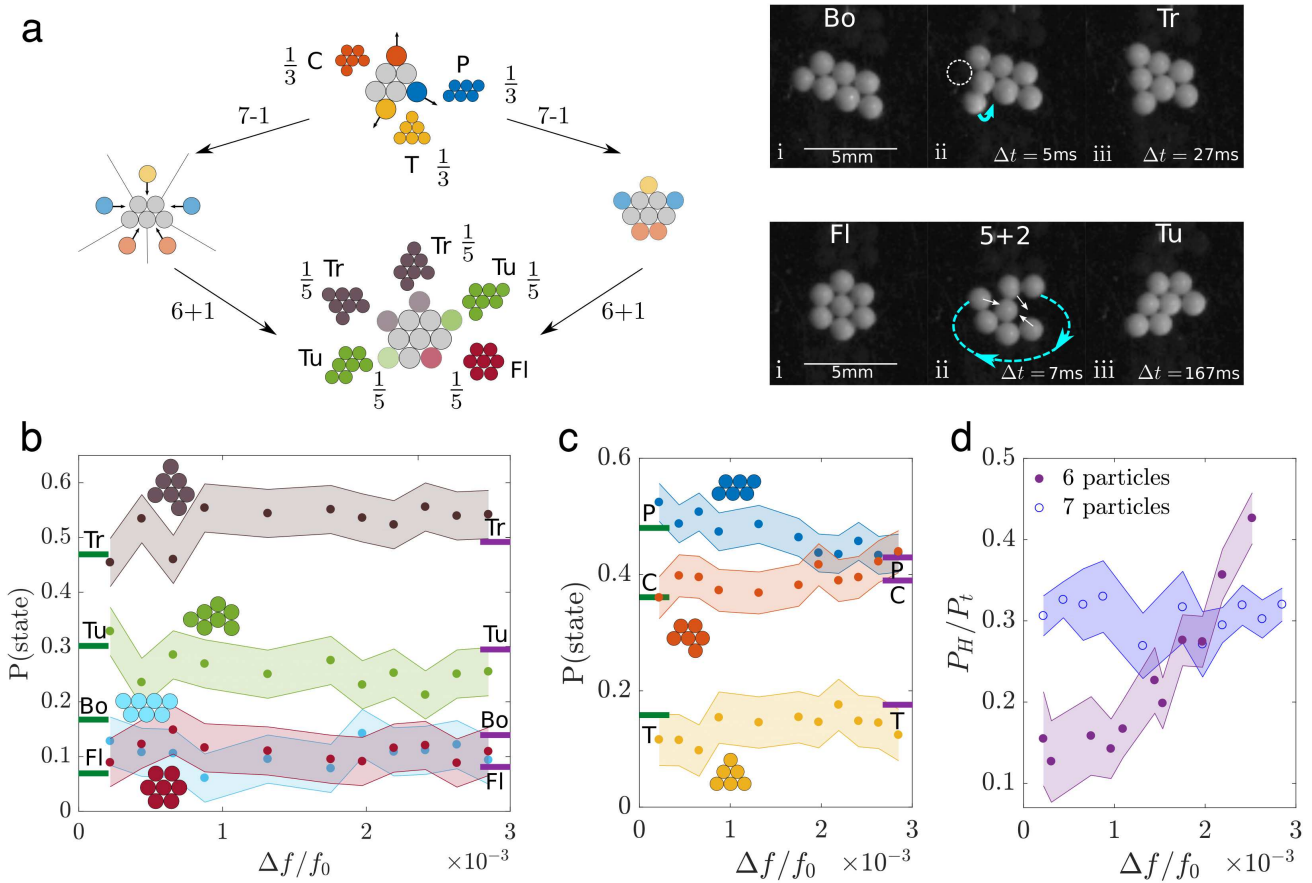}
\caption{\textbf{Seven-particle cluster assembly, ground-state statistics, and transition states.} (a) Left: Schematic of model for statistics of six- and seven-particle clusters in the acoustic trap. Acoustic frequency tunes the formation of six-particle clusters from sticky (centre left) to ergodic (centre right) regimes. Six-particle clusters can be formed when a particle  is removed from the edge of a seven-particle cluster (top). In the reverse process, seven-particle clusters are formed ergodically from six-particle clusters (bottom). Right: Sequence of images from below showing a transition between seven-particle states via a hinge motion (top) or a particle ejection (bottom). See Supplementary Movies 3 and 4 for dynamics. (b) Distribution of seven-particle cluster configurations as a function of detuning parameter, with standard error indicated by shaded area. The green (purple) bars indicate statistics derived from taking into account the sticky (ergodic) six-particle clusters. (c) Statistics of intermediate six-particle clusters within a seven-particle system, plotted as a function of detuning parameter (standard error indicated by shaded area). The green (purple) bars indicate statistics derived from considering sticky (ergodic) cluster assembly. (d) Probability of observing a hinge motion~$P_H$ as a fraction of the total number of transitions~$P_t$ for different values of the detuning parameter. Filled, purple (hollow, blue) disks correspond to clusters of 6 (7) particles.}
\label{fig:seven}
\end{figure*}

Having obtained $\mathrm{T}_6$ and $\mathrm{Q}_5$, we can now make predictions for the case that there are seven particles in the trap and $\mathbf{P}_7$ represents the four ground states shown in Fig.~\ref{fig:setup}d.  Figure~\ref{fig:seven}a shows the reconfiguration pathways for 7-particle clusters and, as examples, transitions from boat to tree via hinge-motion and from flower to turtle via particle ejection and recapture. In the model, we assume that $\mathrm{T}_7$ contains only processes that generate 7- from 6-particle states in an ergodic fashion. As a result, $\mathrm{T}_7$ is a $4\times 3$ matrix with elements corresponding to docking one particle at any available 6-cluster site with equal probability (Fig.~\ref{fig:seven}a).

Recursively solving Eq.~(\ref{eq:general}) for $\mathbf{P}_7$, we find steady-state probabilities near 0.075, 0.47, 0.30 and 0.15 for the flower (Fl), tree (Tr), turtle (Tu) and boat (Bo) configurations (see Methods for details). 
Importantly, the model indicates that all four 7-particle ground states should be largely insensitive to whether the 6-particle intermediate states are formed from 5-particle precursors via a sticky or ergodic process. 
These numerical values are in excellent agreement with the data (Fig.~\ref{fig:seven}b). 

A further model prediction concerns the probabilities for the intermediate 6-particle states in the 7-particle system, shown in Fig.~\ref{fig:seven}c. 
As before, these states are strongly affected by whether the sticky or ergodic assembly process is followed.  
However, the probabilities differ from those for the ground states in the 6-particle system (Fig.~\ref{fig:statistics}b), since now  $\mathrm{T}_7$ and $\mathrm{Q}_6$ enter the Markov-chain model. 
Again we find that these probabilities are consistent with the data.

This match between model and experiments justifies, \textit{a posteriori}, the above assumption about the applicability of the ergodic form of $\mathrm{T}_7$ across the whole range of ~$\Delta f/f_0$. 
However, we can also check this assumption directly.
This is done in Fig.~\ref{fig:seven}d, where we plot the experimentally observed probability of reconfiguration via hinge motion~$P_H$ relative to $P_T$ as a function of the detuning parameter~$\Delta f/f_0$.
While for 6-clusters this fraction increases steadily with detuning, for 7-clusters it is effectively independent of~$\Delta f/f_0$, just as the 7-cluster statistics. 
This difference in hinge-mode proliferation reflects that larger clusters support more bending modes and generate larger out-of-plane bending amplitudes along their periphery. 
We conclude that hinge motions serve as a key indicator for processes that generate ergodic reconfigurations among the ground states. 

In this paper we introduced acoustic levitation as a tool to assemble small clusters of particles and investigate their dynamic reconfigurations. 
While thermal fluctuations set the magnitude of depletion forces in more microscopic particle systems such as colloids, the sonic depletion forces discussed here depend on the sound pressure amplitude and thus can be controlled independently from the effective temperature of the system.
The effective temperature, in turn, was shown to be an emergent quantity that arises from the dynamic response of the levitated objects in the off-resonant regime $\Delta f/f_0 >$ 0.

We can envision acoustic levitation as a more general platform for non-invasive manipulation of granular matter with tunable attractive interactions.   
Our results open up new opportunities for investigating in the underdamped regime the dynamics of extended, 2D rafts of close-packed particles~\cite{Jambon2017}. 
Since the levitated particles are macroscopic, anisotropy in sonic depletion forces could be achieved via particle shape and/or by combining materials with different sound scattering properties.
This may provide a means to assemble complex structures similar to what has been done with patchy colloids~\cite{Chen2011,Kraft2012} or shape-dependent entropic forces~\cite{vanAnders2013}.
Longer-range interactions analogous to those between particles at curved fluid interfaces~\cite{Cavallaro2011} could be implemented using the back-action of levitated grains on the sound field itself. 

\textbf{Acknowledgments} We thank S. Waitukaitis, N. Schade, T. Witten, S. Nagel and R. Behringer for insightful discussions, and J. Z. Kim for a critical reading of the manuscript. We dedicate this work to the memory of R. Behringer. The research was supported by the National Science Foundation through grants DMR-1309611 and DMR-1810390. A.S. and V. V. acknowledge primary support through the Chicago MRSEC, funded by the NSF through grant DMR-1420709.

\textbf{Author Contributions} M.X.L. and H.M.J. conceived of the project and designed the experiments. M.X.L. performed the experiments and analysed the data. M.X.L and A.S. calculated the sonic depletion forces. M.X.L., A.S. and V.V. developed the model and performed the theoretical analysis. All authors contributed to writing the manuscript.

\pagebreak

\include{methods}

\end{document}

%% file: methods.tex
\clearpage
\setcounter{equation}{0}
\renewcommand{\theequation}{S\arabic{equation}}
\subsection*{Methods}
\textbf{Experiment and Data Analysis.}
We used a commercial transducer (Hesentec Rank E) to generate ultrasound. An aluminium horn was bolted onto the transducer to maximise the strength of the nodes in the pressure field, following the finite-element optimisation reported in Ref.~\cite{Andrade10}. The base of the horn (diameter 38.1mm) was painted black to better image the particles from below. The transducer was driven by applying an AC peak-to-peak voltage of 180V, produced by a function generator (BK Precision 4052) connected to a high-voltage amplifier (A-301 HV amplifier, AA Lab Systems). The acrylic reflector was mounted on a lab jack and adjusted to a transducer- reflector distance ~$\lambda_0$, corresponding to~$f_0=45.651$kHz. The acoustic trap was detuned by adjusting the frequency ~$f$ of the function generator. 

As particles we used polyethylene spheres (Cospheric, material density~$\rho = 1,000\:\mathrm{kg}\: \mathrm{m}^{-3}$, diameter~$d=710-850\: \mu \mathrm{m}$). The particles were stored and all experiments were performed in a humidity- and temperature-controlled environment (40-50\% relative humidity, 22-24$\degree$ C). The acrylic reflector was cleaned with compressed air, ethanol and de-ionised water before each experiment. We neutralised any charges that remained on the reflector with an anti-static device (Zerostat 3, Milty).

%We visualise the clustering dynamics of levitated particles in the pressure node parallel to the transducer surface by constructing a transparent acrylic reflector, then recording the motion via a mirror angled at 45$\degree$ to the acrylic surface.  Separate experiments were performed where the cluster dynamics were recorded from the side, without use of the mirror, as schematically illustrated in Fig. 1 of the main text. This setup is front lit with a halogen lamp. 

For each experimental run six or seven particles were inserted into the trap using a pair of tweezers. Video was recorded using a high speed camera (Phantom v12) at 1,000 frames per second. 
%The resulting dynamics are recorded until particles leave the trap due to excessively energetic collisions. 

In order to extract cluster shape information from the raw videos, we thresholded the images, then computed properties of the largest connected region in the resulting image using black-and-white
image operations ({\tt regionprops}). These functions are
available in Matlab. Since each cluster is associated with a specific set of shape parameters, we computed the number of times a cluster shape was formed, divided by the total number of times that any cluster shape was formed, to obtain the cluster statistics in Figs. 2 and 4 of the main text. Hinge motions were similarly obtained (Fig. 4 of main text). 

We calculated the second moment~$J$ of the vertical coordinate~$z$ by integrating the distance to the~$z$ geometric center of the cluster over the area of the cluster. That is, 

\begin{align*}
J=\iint_A (z-z_0)^2 \,dA\, ,
\end{align*}
where~$z_0$ is the~$z$ geometric center of the cluster. Note that we define~$J$ for the specific 2D projection of the cluster sideview.~$J$ is then computed similarly to the cluster topologies and hinge modes from the raw data.

\textbf{Acoustic Force Modeling.}
We used finite element modeling software (COMSOL) to model the force between a pair of particles levitated in the acoustic field, using two different methods. In both cases, we established a background standing pressure wave with given amplitude, with a particle with radius~$r=0.1\lambda$ fixed in the center of the trap. The levitation chamber was constructed to have height~$3\lambda_0/2$ and width~$4\lambda_0$, with materials parameters chosen to match the experiment. In one case, labelled ``point particle" in Fig. 1e of the main text, we computed the force on a point particle in the resulting pressure field by solving the equations for the acoustic field by using the expression derived in Ref.~\cite{Gorkov1961}. In the second case, labelled ``fluid-structure interaction" in Fig. 1e of the main text, we computed the force on a second particle of radius~$r=0.1\lambda$ by computing the full fluid-structure interaction, following the method of Ref.~\cite{GlynneJones2013}. 

\textbf{Markov-Chain Model.}
We consider a discrete-time Markov chain that relates the cluster statistics for five-, six-, and seven-particle clusters by examining the physical processes that produce different clusters. We consider the following mechanisms: (1) Seven particle clusters are formed by ergodically adding a particle to a six-particle cluster (meaning that the particle occupies any binding site with equal probability). (2) Six-particle clusters are formed from five-particle clusters, in a way that depends on the detuning parameter. (3) Six-particle clusters are also formed from the removal of a particle from the edge of a seven-particle cluster. (4) Five-particle clusters are formed from the removal of a particle from the edge of a six-particle cluster.  
Denoting the probability of state~$S$ as~$P(S)$, we write

\begin{align*}
\mathbf{P_7}=\begin{pmatrix}
P(Fl)\\P(Tu)\\P(Tr)\\P(Bo)
\end{pmatrix}, 
\mathbf{P_6}=\begin{pmatrix}
P(P)\\P(C)\\P(T)
\end{pmatrix},
\mathbf{P_5}=\begin{pmatrix}
P(5)
\end{pmatrix}\, .
\end{align*}

We recall that there are four possible states for seven-particle clusters, three for six-particle clusters, and one for five-particle clusters. Let~$\mathrm{T}^{e,s}_{N}$ denote the creation matrix that describes building a~$N$-cluster from an~$(N-1)$-cluster for either ergodic or sticky processes, and ~$\mathrm{Q}_{N}$ the destruction matrix for breaking an~$N+1$-cluster to make a~$N$-cluster. Then 

\begin{align}
\mathbf{P_7}&=\mathrm{T}^e_{7}\mathbf{P_6}, \label{eq:p77}\\
\mathbf{P_6}&= \frac{1}{2} \mathrm{Q}_{6}\mathbf{P_7}+\frac{1}{2} \mathrm{T}^{e,s}_{6}\mathbf{P_5}, \label{eq:p76}\\
\mathbf{P_5}&= \mathrm{Q}_{5}\mathbf{P_6}.
\label{eq:p75}
\end{align} 
Note that we assign equal weight to the processes which form a six-particle cluster from a five-cluster, and those which form a six- cluster from a seven-cluster. In addition,~$T_{6}$ describes either ergodic or sticky six-particle formation processes depending on the detuning parameter.

\subsubsection*{Six particle statistics}

If we exclude the seven-particle processes from the model, we are left with 
\begin{align}
\mathbf{P_6}&= \mathrm{T}^{e,s}_{6}\mathbf{P_5},\label{eq:p66}\\
\mathbf{P_5}&= \mathrm{Q}_{5}\mathbf{P_6}.
\label{eq:p65}
\end{align} 
We construct an effective transition matrix~$R_{66}$, describing the six- to six-cluster transitions through intermediate five-cluster states. Substituting Eq.~(\ref{eq:p65}) into Eq.~(\ref{eq:p66}), 
\begin{align}
\mathrm R_{66} = \mathrm{T}^{e,s}_{6}\mathrm{Q}_{5}.
\end{align}
To find~$\mathrm{Q}_{5}$, we consider the possible clusters that result from removing a particle from the edge of a cluster. Trivially, removing any particle from a six-cluster results in the unique five-cluster: 
\begin{align}
\mathrm{Q}_{5}&= \begin{pmatrix}
1&1&1
\end{pmatrix}.
\label{eq:Q65}
\end{align}
In addition,~$\mathrm{T}^{e,s}_{6}$ are constructed from the ergodic and sticky models: 

\begin{align}
\mathrm{T}^e_{6}=\begin{pmatrix}
2/5\\
2/5\\
1/5
\end{pmatrix}, 
\mathrm{T}^s_{6}=\begin{pmatrix}
1/2\\
1/3\\
1/6
\end{pmatrix}.
\label{eq:T65}
\end{align}

Since the steady state probability vector~$\mathbf{P_6}$ satisfies~$\mathbf{P_6} = R_{66}\mathbf{P_6}$, we find~$\mathbf{P_6}$ by finding the eigenvector of~$R_{66}$ with unit eigenvalue. Substituting 
\begin{align*}
\mathbf{P^e_{6}}=\begin{pmatrix}
2/5\\
2/5\\
1/5
\end{pmatrix}, 
\mathbf{P^s_{6}}=\begin{pmatrix}
1/2\\
1/3\\
1/6
\end{pmatrix} \, .
\end{align*}
These probabilities are shown in Fig. 2b of the main text. 

\subsubsection*{Seven particle statistics}

Similarly to the six-cluster derivation, we derive expressions for the effective transition matrices~$\mathrm{M}_{77}$ and~$\mathrm{M}_{66}$ from Eqs.~(\ref{eq:p77})-(\ref{eq:p75}), such that~$ \mathbf{P_7}=\mathrm{M}_{77}\mathbf{P_7}$ and~$ \mathbf{P_6}=\mathrm{M}_{66}\mathbf{P_6}$. The steady-state probabilities are then the eigenvectors of~$\mathrm{M}_{66}$ and $\mathrm{M}_{77}$ with unit eigenvalue. Note that~$\mathrm{M}_{77}$ and~$\mathrm{M}_{66}$ include transitions through five- and six-cluster intermediates.  Substituting Eqs.~(\ref{eq:p77}) and~(\ref{eq:p75}) into~(\ref{eq:p76}), we obtain

\begin{align}
M_{66}&= \frac{1}{2}\mathrm Q_{6}\mathrm T^e_{7} +\frac{1}{2}\mathrm T^{e,s}_{6}\mathrm Q_{5}\, .
\label{eq:eigensystem6}
\end{align}

We derive~$\mathrm M_{77}$ by substituting Eq.~(\ref{eq:p75}) into (\ref{eq:p76}), which is then substituted for~$\mathbf{P_6}$ in Eq.~(\ref{eq:p77}):

\begin{align*} 
\mathbf{P_7}&=\mathrm T^e_{7}\left( \frac{1}{2} \mathrm Q_{6}\mathbf{P_7}+\frac{1}{2} \mathrm T^{e,s}_{6} \mathrm  Q_{5}\mathbf{P_6} \right)\\
&= \frac{1}{2} \mathrm T^e_{7}\mathrm Q_{6}\mathbf{P_7}+\frac{1}{2}\mathrm T^e_{7} \mathrm T^{e,s}_{6} \mathrm Q_{5}\mathbf{P_6}
\end{align*}

In order to get a closed-form expression for~$\mathbf{P_7}$, we continue substituting for~$P_6$: 

\begin{align*} 
\mathbf{P_7}&= \frac{1}{2} \mathrm T^e_{7}\mathrm Q_{6}\mathbf{P_7}+\frac{1}{2}\mathrm T^e_{7} \mathrm T^{e,s}_{6} \mathrm Q_{5}\left( \frac{1}{2} \mathrm Q_{6}\mathbf{P_7}+\frac{1}{2} \mathrm T^{e,s}_{6} \mathrm Q_{5}\mathbf{P_6} \right)\\
&=\frac{1}{2} \mathrm T^e_{7}\mathrm Q_{6}\mathbf{P_7}+\frac{1}{4}\mathrm T^e_{7} \mathrm T^{e,s}_{6} \mathrm Q_{5}\mathrm Q_{6}\mathbf{P_7}\\
&+\frac{1}{4}\mathrm T^e_{7} \mathrm T^{e,s}_{6} \mathrm Q_{5} \mathrm T^{e,s}_{6} \mathrm Q_{5}\mathbf{P_6} 
\end{align*}

This leads to a geometric series in increasing numbers of transitions between five- and six-cluster states:
 
\begin{align*} 
\mathbf{P_7}&=\frac{1}{2} \mathrm T^e_{7}\mathrm Q_{6}\mathbf{P_7}+\left( \sum_{n=1}^{\infty}\frac{1}{2^{n+1}}\mathrm T^e_{7} (\mathrm T^{e,s}_{6} \mathrm Q_{5})^n \mathrm Q_{6}\right) \mathbf{P_7} 
\end{align*}

We note that~$\mathrm T^{e,s}_{6} \mathrm Q_{5}$ is idempotent, so that $(\mathrm T^{e,s}_{6} \mathrm Q_{5})^n = \mathrm T^{e,s}_{6} \mathrm Q_{5}$ for any~$n$. Then we complete the geometric series and write 

\begin{align}
M_{77}&= \frac{1}{2}\mathrm T^e_{7}\mathrm Q_{6} +\frac{1}{2}\mathrm T^e_{7}\mathrm T^{e,s}_{6}\mathrm Q_{5}\mathrm Q_{6}
\label{eq:eigensystem7}
\end{align}

To find the destruction matrix~$\mathrm Q_{6}$, we assume that any particle on the edge of a cluster has equal probability to be removed. Then Fl can only make C, Tu makes P and C with equal probability, Tr makes P, C, and T equally, and Bo makes only P:

\begin{align*}
\mathrm Q_{6}&=\begin{pmatrix}
0&1/3&1/2&1\\
1&1/3&1/2&0\\
0&1/3&0&0
\end{pmatrix}
\end{align*}

Similarly, we construct~$\mathrm T^e_{7}$ assuming that a seventh particle has equal probability to attach to any binding site on a six-particle cluster: 

\begin{align*}
\mathrm T^e_{7}&=\begin{pmatrix}
0&1/5&0\\
1/3&2/5&1\\
1/3&2/5&0\\
1/3&0&0
\end{pmatrix}
\end{align*}

Substituting into Eqs.~\ref{eq:eigensystem6} and~\ref{eq:eigensystem7} and solving the eigenvalue problem, as for the six-particle clusters, gives

\begin{align*}
\mathbf{P^s_7}=\begin{pmatrix}
0.071\\
0.464\\
0.303\\
0.161
\end{pmatrix},
\mathbf{P^e_7}=\begin{pmatrix}
0.079\\
0.480\\
0.299\\
0.141
\end{pmatrix}
\end{align*}

and

\begin{align*}
\mathbf{P^s_6}=\begin{pmatrix}
0.484\\
0.355\\
0.161
\end{pmatrix},
\mathbf{P^e_6}=\begin{pmatrix}
0.426\\
0.349\\
0.180
\end{pmatrix}
\end{align*}